\documentstyle[pre,aps,epsfig,twocolumn]{revtex}

\begin{document}

\title{Non-Monotonic Dispersion of Surface Waves in Magnetic Fluids}

\date{J.~Magn.~Magn.~Materials: received Dec 20, 1995, accepted Mar 18, 1996}

\author{Thomas Mahr}
\address{ Institut f\"ur Experimentelle Physik,
          Otto-von-Guericke-Universit\"at,
          Postfach 4120,
          D-39016 Magdeburg,
          Germany}

\author{Alexander Groisman}
\address{ Department of Physics of Complex Systems,
          Weizmann Institut,
          76100 Rehovot,
          Israel}

\author{Ingo Rehberg}
\address{ Institut f\"ur Experimentelle Physik,
          Otto-von-Guericke-Universit\"at,
          Postfach 4120,
          D-39016 Magdeburg,
          Germany}

\maketitle

\begin{abstract}

The dispersion relation of surface waves of a magnetic fluid in a
magnetic field is studied experimentally. We verify the
theoretically predicted existence of a non-monotonic dispersion
relation. In particular, we demonstrate the existence of two
different wave numbers occuring at the same frequency in an annular geometry.

\end{abstract}


\section{Introduction}

The science of magnetic fluids is a multidisciplinary field, which
encompasses physics, chemistry, mathematics, engineering and medical
sciences \cite{jmmm95}. One of its interesting features is the
spontaneous formation of patterns under the influence of an external
magnetic field \cite{RosenBuch}. If a static magnetic field is
applied normal to the surface, the so called normal
field- or Rosensweig-instability will occur at a critical value of
the magnetisation leading to a deformation of the surface.
Due to theoretical considerations \cite{RosenBuch}, the advent of
this pattern forming instability is accompanied by a non-monotonic
dispersion relation. This interesting behaviour is not a unique
feature of magnetic fluids, however. Similar behaviour is expected
for polarizable fluids under the influence of a normal electric field
\cite{Taylor}. Both theories apply for an inviscous fluid in an
infinite layer.
Although experimental studies of externally induced surface waves
have been published \cite{Zelazo,Bashtovoi},
an experimental verification of the non-monotonic
dispersion of surface waves is missing so far.
We thus present an experimental attempt to verify the existence of
a non-monotonic dispersion relation for surface waves in a magnetic
fluid.

\section{Experimental Setup and Procedure}

\begin{figure}[th]
\epsfig{file=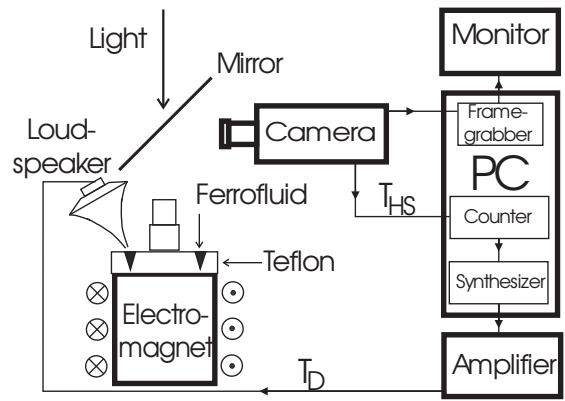,width=8cm}
\caption {Experimental setup. }
\label{aufbau}
\end{figure}

The experimental setup is shown in Fig.~\ref{aufbau}. We use the
commercially available magnetic fluid EMG 909 (Ferrofluidics) with
the following properties:
density $\rho = 1020$ kgm$^{-3}$,
surface tension $\sigma = 2.65 \cdot 10^{-2}$ kgs$^{-2}$,
initial magnetic permeability $\mu = 1.8$,
magnetic saturation $ M_S = 1.6 \cdot 10^4$ Am$^{-1}$,
viscosity $ \eta= 6 \cdot 10^{-3}$ Nsm$^{-2}$.

\begin{figure}[th]
\epsfig{file=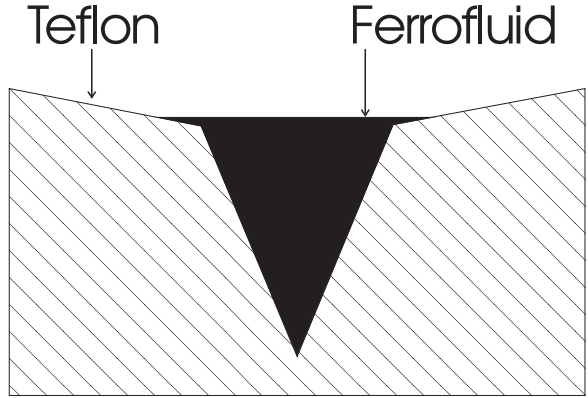,width=8cm}
\caption {Cross-section of the V-shaped teflon channel.}
\label{channel}
\end{figure}

The fluid is filled into a V-shaped \cite{BacriSalin} cylindrical
teflon channel of 6 cm diameter as indicated in Fig.~\ref{channel}.
The channel has a depth of 6 mm and upper width of 5 mm. The
motivation for using the V-shape is to minimize field-inhomogeneities
at the boundaries of the channel.
The upper part of the channel has a slope of $15^{\circ}$, which is
close to the measured contact angle between EMG 909 and teflon, in
order to enforce a flat surface of the fluid.

The channel is placed on the top of an electromagnet, with an iron
core of 9 cm diameter and 15 cm height. The coil consists of 250
windings of copper wire with a diameter of 2 mm. A current of 4 A
is then sufficient to produce the magnetic field used in these
experiments. The two additional iron cylinders indicated in
Fig.~\ref{aufbau} on top of the teflon piece have a diameter 3.5 cm
and height 3 cm, and diameter 3 cm and height 3 cm, respectively.
They serve to increase the homogeneity of the magnetic field near
the surface of the fluid.
The field is monitored by means of a hall probe
(Koshava 3 Teslameter) located near the surface of the channel.

In order to excite surface waves without mechanical contact with
the magnetic fluid,
we use an air jet created by a loudspeaker of diameter 10 cm,
which is directed towards the surface of
the fluid with the help of a nozzle as indicated in
Fig.~\ref{aufbau}.
The loudspeaker is driven by a sinusoidal current with
computer-controlled amplitude and frequency.

The shadowgraph method \cite{shadow} is used for detecting the
surface waves. The light from a tungsten bulb, placed 50 cm above
the center of the channel,
is reflected at the surface of the fluid and directed towards the
CCD-camera with a glass plate acting as a semitransparent mirror
as indicated in Fig.~\ref{aufbau}.
In order to avoid strong nonlinearities of the optical response,
the lens of the camera focuses on a plane only a few millimeters
above the surface of the fluid. A more precise determination of
the focal plane is not necessary because no attempt is made to
measure the amplitude of the surface deformation, as this is not
considered as an important parameter for these experiments.

The analysis of the images and the control of the experiment is done
with a 66 MHz 80486-PC, equipped with a $512 \times 512$ 8 bit frame
grabber (Data Translation DT2853), a programmable counter (8253)
located on a multifunction I/O-board (Meilhaus ME-30), and a
synthesizer-board (STAC ASP-ADW1).

The CCD-camera (Phillips LDH 0600/00) works in the
interlaced mode at 50 Hz using an exposure time of 40 ms.
The time between two horizontal synchronization pulses is
$T_{HS} = 64$ $\mu$s.
The corresponding frequency of $1/T_{HS} = 15625$ Hz is divided by
the programmable counter.
The reduced frequency at the output of the counter acts as a pacer
for the synthesizer-board, driving the loudspeaker via an amplifier
(Euro Test LAB/S 115).

The camera takes images at a fixed frequency of 25 Hz, while the
driving frequency for the surface waves of interest reaches up to
30 Hz. The ensuing problem in resolving the dynamics is overcome by a
software-realized stroboscopic image acquisition, which is based on
the computer-controlled synchronization between driving and sample
frequency.

The stroboscopic image acquisition works in the following way:
The period $T_D$ of the driving oscillation can only be an integer
multiple of the time $T_{HS}$ between horizontal synchronization
pulses according to
\[
T_D = N_{SB} M_C T_{HS},
\]
where $M_C$ is the multiplier value loaded in the counter,
and $N_{SB}$ is the number of data points used by the
synthesizer-board to represent the sine wave. The images can only
be sampled with a time interval $T_S = 625$~$T_{HS} = 40$ ms,
which is the time between two full frames.
Consecutive images are taken at times $t_j = j T_S$,
where $j$ represents the number of the image.
If one assumes the response of the surface to be periodic with the
driving period $T_D$, then the surface at time $t_j$ is in the same
state as at time $t'_j < T_D$, according to:
\begin{eqnarray*}
t'_j(t_j) & = & t_j {\rm mod} T_D = (j T_S) {\rm mod} T_D \\
          & = & ((625 j) {\rm mod} (N_{SB} M_C)) T_{HS}.
\end{eqnarray*}
This equation describes the projection of the images taken at the
times $t_j$ into the time interval $[0,T_D[$.
The ordering of these images according to increasing $t'_j(t_j)$
yields to the reconstruction of one period of the surface motion.
The total measurement time $T_M$ is the least common multiple of the
two periods $T_D$ and $T_S$. The number of different values of $t'_j$
is $N_I = T_M / T_S$.
The driving frequencies are chosen as a compromise between the desired
frequency and sufficient temporary resolution corresponding to
$N_I \approx 100$.

\begin{figure}[th]
\epsfig{file=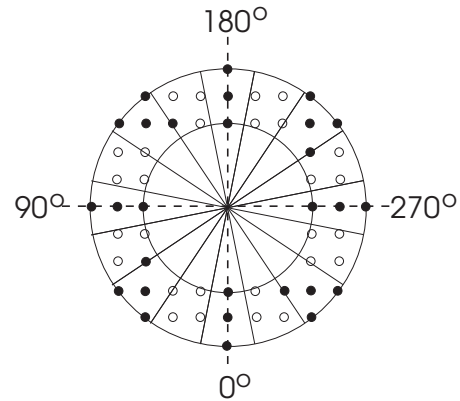,width=8cm}
\caption {The rectangular pixels of the frame camera are
                       assigned to segments along an annulus. Only 16
                       segments are shown here for clarity; for the
                       measurements we used 128 segments. Note that
                       the number of pixels per segment is not
                       constant.}
\label{ring}
\end{figure}

In order to measure the light intensity along the annulus by means of
the frame camera, we divide the circle into 128 segments as
schematically indicated in Fig.~\ref{ring}. The mean value of the
intensities of those pixels which are located within one segment
of the ring is calculated, thus yielding to 128 intensity measurements
along the annulus.
We consider only odd-numbered lines, which correspond to one
half-frame of the camera working in the interlaced mode.
The calculations can be performed with the
video frequency of 25 frames per second by using a look-up-table
technique for the addressing of the pixels of a segment.

\section{Experimental Results}

\begin{figure}[th]
\caption {Space-time-plot for $B=0$,
          and $T_D = 768$ $T_{HS} = 49.152$ ms.}
\label{ort1}
\end{figure}
\begin{figure}[th]
\epsfig{file=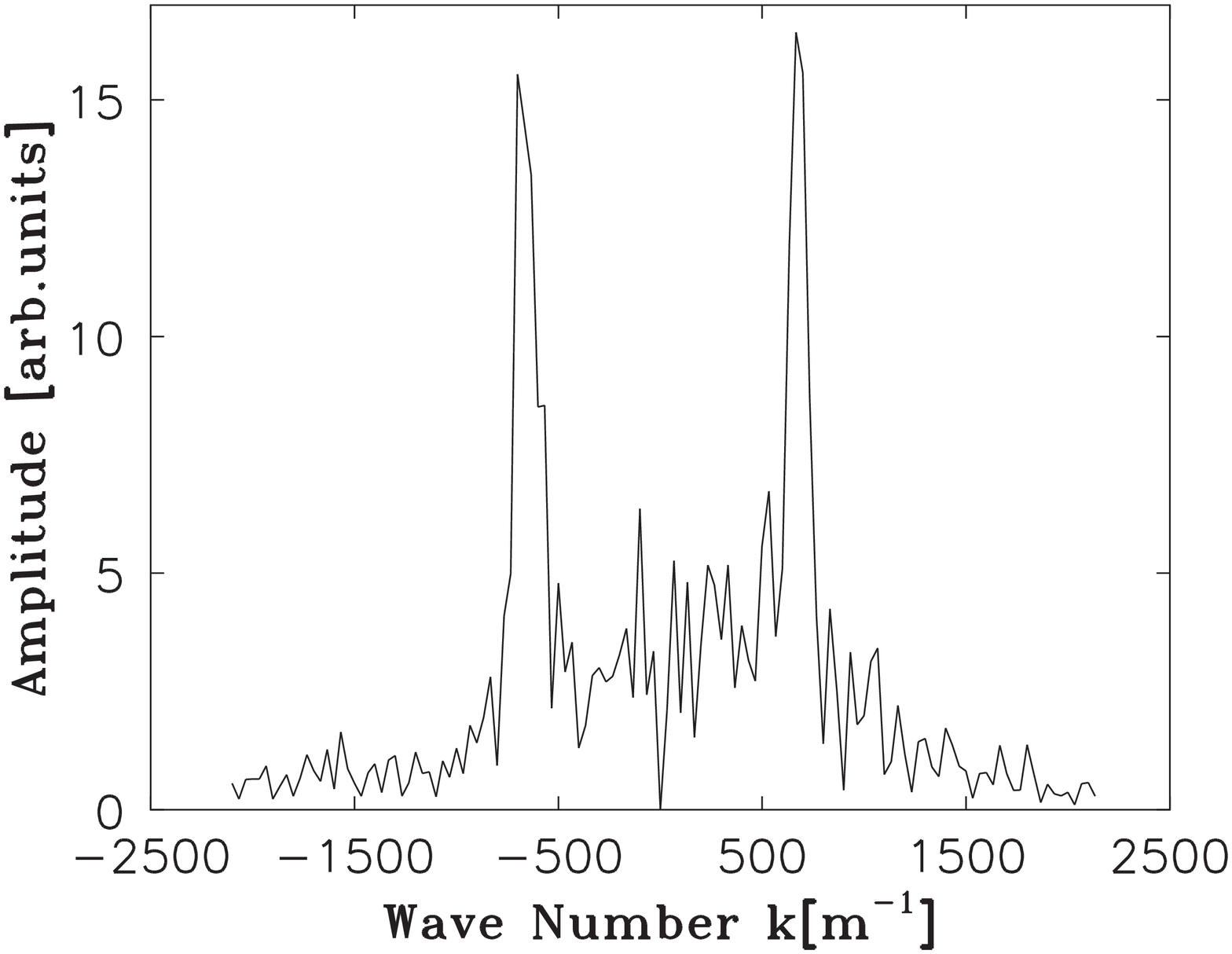,width=8cm}
\caption {Spectrum obtained from Fig.~\ref{ort1}.
          Only wave numbers corresponding to a frequency of
          $T_D^{-1} = 20.345$ Hz are shown.}
\label{fft1}
\end{figure}
Fig.~\ref{ort1} demonstrates the behavior of a surface wave
which is driven at an excitation period $T_D=49.152$ ms
in the absence of a magnetic field.
The horizontal axis represents the 128 intensity values measured
along the annulus. The driving occurs at $0^{\circ}$ and
$360^{\circ}$, respectively.
In this case the stroboscopic algorithm yields 256 measurements per
period.
These 256 lines are plotted on top of each other according to the
sorting algorithm described above. The image displays two waves
emanating from the source, travelling clockwise and counterclockwise
along the channel.
The small amplitude at the position of $180^{\circ}$ results mainly
from the attenuation and presumably partly from the interference of
the two waves.
In order to extract the wave numbers we perform a two-dimensional FFT
\cite{NumRec} of this image.
In this case this method is artifact-free because we have periodic
boundary conditions both in space and in time. The FFT leads to a
two dimensional complex field $H(k,\omega)$. The frequency
of interest is $\omega_1 = 2 \pi / T_D$.
In Fig.~\ref{fft1} we have plotted $|H(k,\omega_1)|$ as a function
of the 128 available wave numbers. Negative wave numbers represent
the wave travelling in clockwise direction, while positive wave
numbers correspond to the ones travelling counterclockwise. The
spectrum clearly shows two peaks indicating the two dominant waves.
The finite size of the peaks is due to the attenuation of the waves.
The wave number $k_P$ is interpolated between the  peak value
$|H(k_{max1},\omega_1)|$ and its largest neighbour
$|H(k_{max2},\omega_1)|$ according to
\[
k_P = \frac{k_{max1} |H(k_{max1},\omega_1)|
    + k_{max2} |H(k_{max2},\omega_1)|}
    {|H(k_{max1},\omega_1)| + |H(k_{max2},\omega_1)|}.
\]
Measuring $k_P$ for different values of the driving period $T_D$
then yields to the dispersion relation (see Figs. \ref{disprel2}
and \ref{disprel1} below). In principal this experimental procedure
would also allow to determine the attenuation of surface waves,
provided that our optical setup responds in a linear manner.

\begin{figure}[th]
\epsfig{file=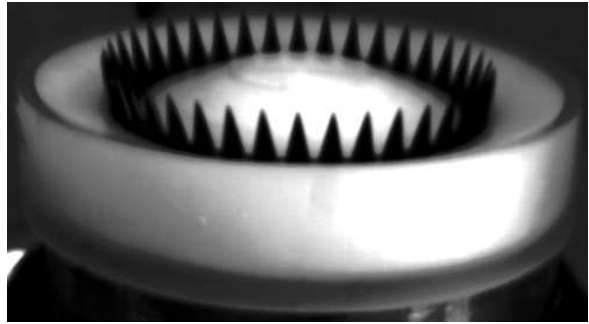,width=8cm}
\caption {Fully developed peaks of a magnetic
          fluid obtained at $B > B_c$ in the annular channel.}
\label{rosentorte}
\end{figure}
If a sufficiently high magnetic field perpendicular to the surface
of the magnetic fluid is applied, the normal field instability occurs,
leading to characteristic peaks shown in Fig.~\ref{rosentorte}.
Because of problems with the aging of the magnetic fluid, the critical
magnetic field $B_c$ for the onset of this instability drifts in a
few weeks from $B_c = 0.009$ T to $0.011$ T.
We thus present all values of the magnetic field in units of the
corresponding critical magnetic field.

On the basis of theoretical considerations for the onset of the
Rosensweig-Instability \cite{RosenBuch} one can show that
non-monotonic dispersion relations are expected for values of the
magnetic field ranging between $0.93$ $B_c$ and $B_c$. This range
is independent of the parameters of the fluid.
An example for a non-trivial behavior of the surface waves in this
regime is shown in Fig.~\ref{ort2} for a value of the magnetic field
of $0.93$ $B_c$. This space-time plot displays a clearly more
complicated behavior than Fig.~\ref{ort1}. It seems to indicate
that more than one wave number is excited by the driving period
$T_D=196.608$ ms. The corresponding spectrum is shown in
Fig.~\ref{fft2}. The extraction of the dominant wave number
is easily obtained, while peaks corresponding to higher wave numbers
are much less pronounced and cannot be obtained from this spectrum
with reasonable accuracy. We have thus decided to extract only one
wave number from the spectra.
The results for counter-clockwise travelling waves are summarized in
Fig.~\ref{disprel2}, and for the clockwise waves in
Fig.~\ref{disprel1}.
The ordinate of the plot corresponds to the driving frequency, while
the abscissa indicates the wave number obtained from the spectra.
The results for zero magnetic fields are shown as crosses.
\begin{figure}[th]
\caption {Space-time-plot for $B=0.93$ $B_c$,
          and $T_D = 3072$ $T_{HS} = 196.608$ ms.}
\label{ort2}
\end{figure}
\begin{figure}[th]
\epsfig{file=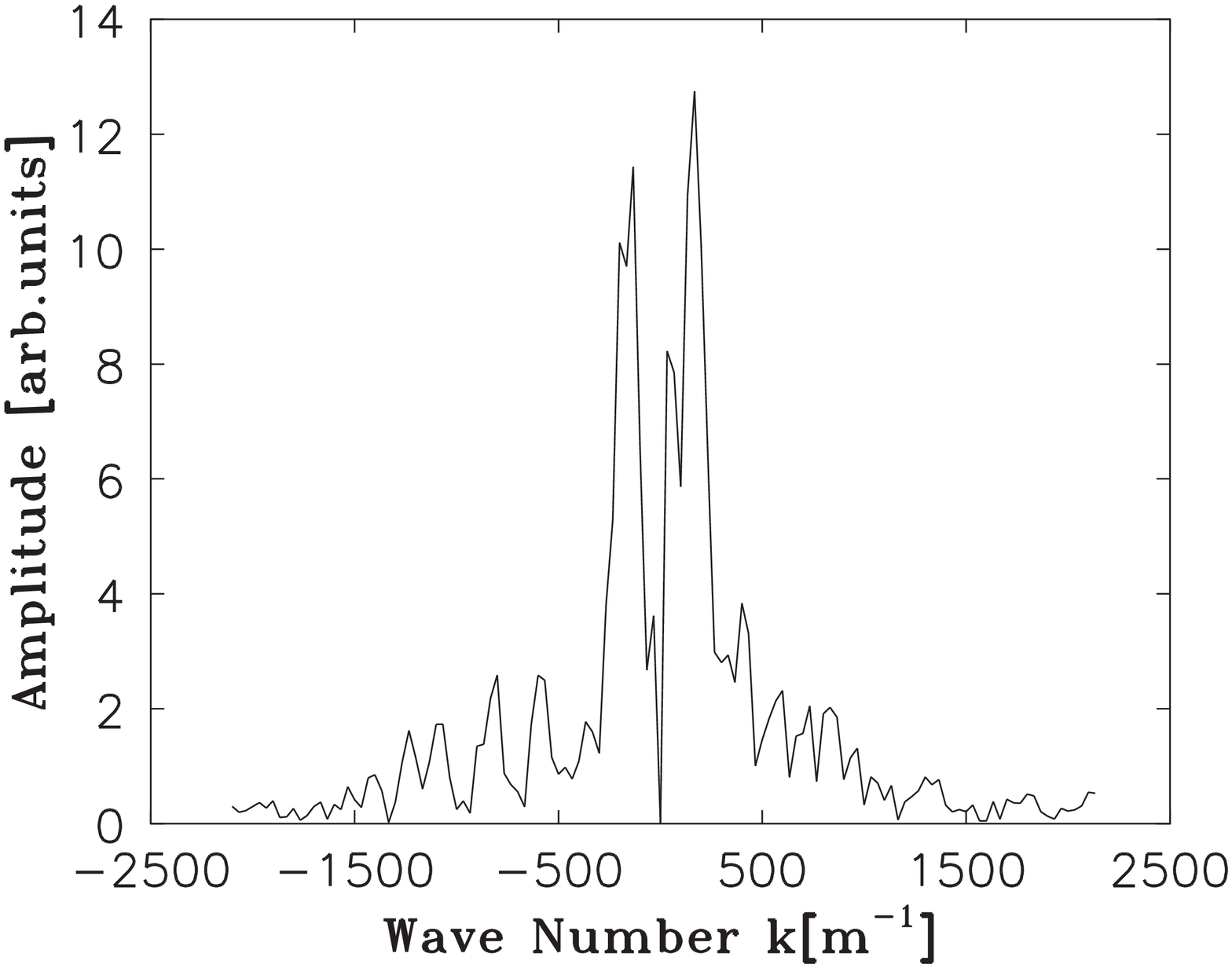,width=8cm}
\caption {Spectrum obtained from Fig.~\ref{ort2}.
          Only wave numbers corresponding to a frequency of
          $T_D^{-1} = 5.09$ Hz are shown.}
\label{fft2}
\end{figure}

The solid lines in Figs.~\ref{disprel2} and \ref{disprel1} are fits
of the function $\omega^2 = ak - bk^2 + ck^3$,
where $a$, $b$ and $c$ are fit parameters. The choice of this
mathematical form for the fitting function is inspired by the
dispersion relation obtained for an inviscid fluid and an
infinite layer \cite{RosenBuch}.
There the parameter $a$ would have the value of $g = 9.81$ m/s$^2$.
The mean value of our fit parameter $a$ is $9.96$ m/s$^2$.
The same theory predicts $b$ to be given by the term
$\mu_0 \mu \frac{(\mu-1)^2}{\mu+1} \frac{1}{\rho} H^2$.
The mean value of $b/H^2$ is $2.9 \cdot 10^{-10}$ $\frac{\rm m^4}{\rm s^2 A^2}$.
When using the permeability $\mu = 1.8$ provided by the manufacturer of
the ferrofluid, one expects $5.1 \cdot 10^{-10}$ $\frac{\rm m^4}{\rm s^2 A^2}$.
The parameter $c$ is theoretically predicted to be
$\sigma/\rho = 2.6 \cdot 10^{-5}$ m$^3$/s$^2$ when using the values
of the surface tension and the density given above. The measured mean
value of $c$ is $2.0 \cdot 10^{-5}$ m$^3$/s$^2$. In spite of this fairly
reasonable agreement it must be stressed that our finite channel is not
designed to measure material parameters. There is a visible disagreement
between the data and the fitted curves, which are based on an idealized
theory. The solid lines are thus basically intended to guide the eye.

\begin{figure}[th]
\epsfig{file=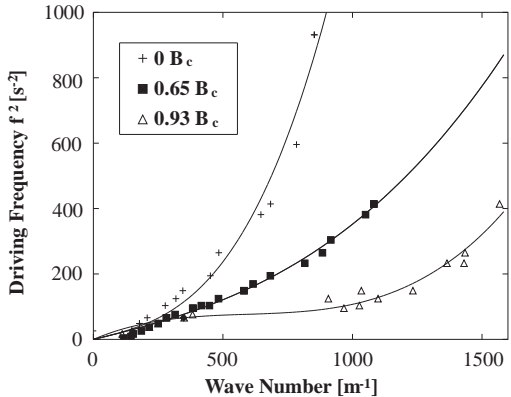,width=8cm}
\caption {Dispersion relation for the waves
          travelling counter-clockwise.}
\label{disprel2}
\end{figure}

\begin{figure}[th]
\epsfig{file=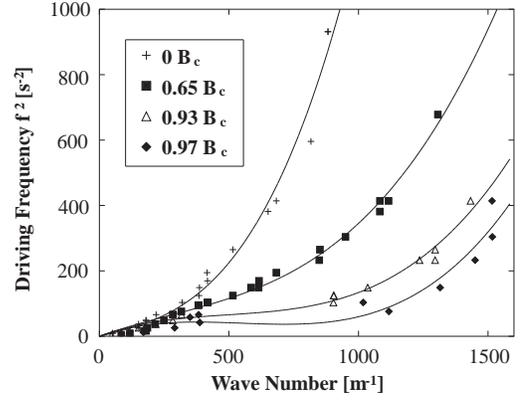,width=8cm}
\caption {Dispersion relation for the waves
          travelling clockwise.}
\label{disprel1}
\end{figure}

In increasing the magnetic field the slope of the dispersion relation
is reduced.
Because of the experimental difficulties described above, we have not
attempted to obtain different wave numbers for one given driving
frequency, but have rather plotted only the wave number $k_P$ of the
largest peak. The prediction of a non-monotonic dispersion relation
is thus not directly verified by the data points presented in the
plot. The fit for $B = 0.97$ $B_c$ in Fig.~\ref{disprel1} shows a
local minimum and makes the interpretation of the data points in
terms of a non-monotonic dispersion relation very likely. Due to
the inhomogeneous magnetic field the corresponding measurement
of the waves travelling counter-clockwise was not possible for this
value of the magnetic field.

\begin{figure}[th]
\caption {Space-time-plot for $B=0.96$ $B_c$,
          and $T_D = 2240$ $T_{HS} = 143.36$ ms.}
\label{ort3}
\end{figure}
\begin{figure}[th]
\epsfig{file=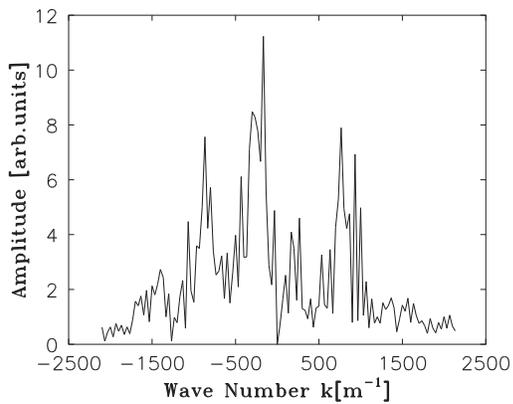,width=8cm}
\caption {Spectrum obtained from Fig.~\ref{ort3}.
          Only wave numbers corresponding to a frequency of
          $T_D^{-1} = 6.98$ Hz are shown.}
\label{fft3}
\end{figure}

In order to demonstrate that more than one wave number is indeed
present in the parameter range of interest, we finally present
in Fig.~\ref{ort3} a measurement for $B = 0.96$ $B_c$ and a
driving period of $T_D=143.36$ ms.
It clearly shows more than one wave number.
In this case the relatively strong attenuation is caused by the
longer driving period (compared to Fig.~\ref{ort1}).
The asymmetry between the left and the right hand side is
believed to be caused by the inhomogeneity of the magnetic field,
which becomes more dominant close to the critical value $B_c$.
The corresponding spectrum is shown in Fig.~\ref{fft3}.
The spectrum might indicate that we reach the limits of our
measurement procedure here. It is fairly complex, presumably partly
due to the effects of a nonlinear optical response, and effects of
field inhomogeneities.
In any case at least two peaks are clearly visible for the wave which
travels clockwise.

\section{Summary and Conclusion}

By using a computer-controlled stroboscope algorithm we have measured
non-monotonic dispersion relations for a magnetic fluid below the
onset of a static pattern forming instability. Due to the finite
geometry of our annular channel a quantitative comparison of the
experimental results with a theoretical calculation is not possible.
Our experiments clearly demonstrate a qualitative similarity
with the theory obtained for a inviscous fluid in an infinite layer.
Thus, the intriguing possibility arises to parametrically excite
\cite{CrossHohenberg,BacriCebers} three different wave numbers with one single
driving frequency, and to study the nonlinear interaction of the
ensuing surface waves.

\section{Acknowledgment}

A.~G. acknowledges support of MINERVA. The experiments are supported
by Deutsche Forschungsgemeinschaft. It is a pleasure to thank
H.~R.~Brand, W.~Pesch, H.~Riecke and J.~Weilepp for stimulating
discussions.

\end{document}